\documentclass[%
 aip,
 amsmath,amssymb,
 reprint,%
]{revtex4-1}

\usepackage[pdftex]{graphicx} 
\usepackage{dcolumn}
\usepackage{bm}

\usepackage[utf8]{inputenc}
\usepackage[T1]{fontenc}
\usepackage{mathptmx}

\usepackage{bm}
\usepackage{bmpsize}
\usepackage{here}
\usepackage{amsmath}
\usepackage{amsfonts}
\usepackage{url}
\usepackage{natbib}
\usepackage{colortbl,array,xcolor}
\usepackage{comment}
\usepackage{bbm}

\usepackage{theorem}
\newtheorem{thm}{Theorem}

{\theorembodyfont{\upshape}
}

\begin{document}

\preprint{AIP/123-QED}

\title[Existence of multiple noise-induced transitions in Lasota-Mackey maps]{Existence of multiple noise-induced transitions in Lasota-Mackey maps}

\author{Takumi Chihara}
\affiliation{%
Department of Mathematics, Hokkaido University, N10 W8, Kita-ku, Sapporo 001-0010, Japan
}%
\author{Yuzuru Sato}%
\email{ysato@math.sci.hokudai.ac.jp}
\affiliation{%
Department of Mathematics, Hokkaido University, N10 W8, Kita-ku, Sapporo 001-0010, Japan
}%
\affiliation{%
RIES, Hokkaido University, N20 W10, Kita-ku, Sapporo 001-0020, Japan
}%
\affiliation{%
London Mathematical Laboratory, 8 Margravine Gardens, London, W6 8RH, UK
}%

\author{Isaia Nisoli}%
\affiliation{ 
Institute de Matem\`{a}tica, Universidade Federal do Rio de Janeiro, Av. Athos da Silveira Ramos 149, Bloco C Cidade Universit\'{a}ria, 68530 21941-909 Rio de Janeiro, Brazil.
}%

\author{Stefano Galatolo}
\affiliation{%
Dipartimento di Matematica, Universit\`{a} di Pisa, Via Buonarroti 1, 56127 Pisa, Italy
}%

\date{\today}

\begin{abstract}
We prove the existence of multiple noise-induced transitions in the Lasota-Mackey map, which is a class of one-dimensional random dynamical system with additive noise. The result is achieved by the help of rigorous computer assisted estimates. We first approximate the stationary distribution of the random dynamical system and then compute certified error intervals for the Lyapunov exponent. 
We find that the sign of the Lyapunov exponent changes at least three times when increasing the noise amplitude. We also show numerical evidence that the standard non-rigorous numerical approximation by finite-time Lyapunov exponent is valid with our model for a sufficiently large number of iterations. Our method is expected to work for a broad class of nonlinear stochastic phenomena.
\end{abstract}

\maketitle

{\bf Noise-induced phenomena emerge in nonlinear dynamics in the presence of noise. The central problem of noise-induced phenomena is to study in which way the asymptotic behavior of the deterministic system is affected by the external noise, and how much its macroscopic behavior is altered. Despite of the simplicity of the problem, most non-trivial noise-induced phenomena have not been analyzed rigorously. Recently, the rigorous computer assisted estimation of statistical properties of random dynamical systems has been developed. We apply these methods to prove the existence of multiple noise-induced transitions in a class of chaotic map with additive noise. }

\section{Introduction}
\label{sec:intro}

Often stochastic noise causes qualitative changes of the statistical and dynamical behaviour of chaotic dynamical systems. For example, a small additive noise can turn a chaotic system into an orderly one, which is called {\it noise-induced order}. Noise-induced order was first discovered in a one-dimensional map constructed from an experimental time series of Belousov-Zhabotinsky reaction \citep{matsumoto1983noise}. Chaos-to-order transitions increasing the noise amplitude were observed through several physical quantities, including the Lyapunov exponent, Kolmogorov-Sinai entropy, and the power spectrum of the dynamics.
Subsequently, noise-induced order was confirmed through measurements 
of experiments of Belousov-Zhabotinsky reaction \citep{yoshimoto2008noise}. Multiple transitions from chaotic regime to orderly regime, and then to a different chaotic regime, then back to a different regular regime, have also been found in models of random dynamical system when we increase the noise amplitude \citep{sato2019noise}. In this paper, we focus on the multiple noise-induced transitions in the Lasota-Mackey map \citep{lasota1987noise} introduced in Section \ref{sec:target}. 

Recently, the existence of noise-induced order in BZ map has been mathematically  proved  by {\it S. Galatolo, et al.}\citep{galatolo2017existence} by validated numerics, showing a change on the sign of the Lyapunov exponent as the noise amplitude increases.
 We apply their methods to the computation of the Lyapunov exponents of the Lasota-Mackey map to show the existence of multiple noise-induced transitions. 


The rigorous approximation of the Lyapunov exponents is based on the approximation of the stationary distribution by the Ulam method \citep{ding2002finite}, which  approximates transfer operators by a finite dimensional transition matrix. 
Note that the Ulam method works specially well for random dynamical systems because the addition of noise simplifies the functional analytic properties of the transfer operators, and smooths out the fine details of the stationary distributions. 

The paper is organised as follows.
In Section \ref{sec:distribution}, we describe the Lyapunov exponent of random dynamical systems and clarify our problem. 
In Section \ref{sec:target}, we introduce Lasota-Mackey maps, a class of random map with additive noise, and discuss noise-induced transitions by non-rigorous numerical estimates, to show the phenomenology and find the right paremeter sets to which apply the computer aided estimates and prove our rigorous results. 
In Section \ref{sec:theory}, we introduce the theoretical background of the rigorous approximation of the Lyapunov exponent, and give bounds for Lyapunov exponents.
In Section \ref{sec:result}, the algorithmic properties of rigorous computation and the final result are shown. 
and we compare the Lyapunov exponent  obtained by the rigorous computation and those obtained by common numerical experiments. 
In Section \ref{sec:conclusion}, we give a summary and an overview.


\section{The Lyapunov exponent}
\label{sec:distribution}
\subsection{Lyapunov exponents of  random dynamical systems}
A one-dimensional random dynamical system with additive noise is given by 
\begin{equation}
\label{determ+noise}
    x_{n+1}=T(x_{n})+\xi_{n}
\end{equation}
where $T:\mathbb{R}\rightarrow \mathbb{R}$ is a piecewise $C^1$ non-singular map, i.e.  a map whose associated pushforward preserves absolutely continuous measures. 
The additive noise term  $(\xi_{i})_{i\in \mathbb{Z}}$ is defined as a series of independent and identically distributed random variables with a probability distribution $\rho_\theta$ having bounded variation and supported in the interval $[-\frac{\theta}{2} , \frac{\theta}{2} ]$ characterising the range of the fluctuation.

Let $\omega:=(\xi_{i})_{i\in\mathbb{Z}}$ be a noise realization. The Lyapunov exponent associated to the point $x_0\in \mathbb{R}$ and the realization $\omega$ of our random dynamics is defined by 
\begin{equation}
    \label{dfn:lyap_infinite_time}
    \lambda_{\theta}(\omega,x_{0})=\lim_{N\rightarrow \infty}\frac{1}{N}\sum_{i=0}^{N-1}\log |T'(x_{i})|,  
\end{equation}
where $x_i$ is defined by the random dynamical system \eqref{determ+noise}. 
The Lyapunov exponent of random dynamical systems characterizes the average expansion rate of orbits as is the case with deterministic dynamical systems. When the random dynamical system has a stationary measure $\mu$ which is ergodic, the Lyapunov exponent is $\mu$-almost surely a constant. 
\begin{equation}
\label{dfn:lyap_finite_time_and_lyap}
     \lambda_{\theta}(\omega,x_{0})=\overline{\lambda_{\theta}.}
\end{equation}

\subsection{Transfer operator and stationary distribution for random dynamical systems}
For the random map \eqref{determ+noise},  
we consider the case of a fixed noise $\xi_n=\xi$; 
fixed a noise we have 
a deterministic transformation; 
\begin{equation}
    T_{\xi}=T(x)+\xi . 
\end{equation}
For such a transformation the transfer operator \footnote{sometimes called the push-forward operator when acting on measures} $L_{\xi,\theta}:L^{1}\rightarrow L^{1}$ for $T_{\xi}$ is defined by the equation 
\begin{equation}
 \int_{A} (L_{\xi,\theta}f)(x) dx=\int_{T_{\xi}^{-1}(A)}f(x)dx.
\end{equation}
where $f\in L^{1}$ and $A$ is a Borel measurable set.
Statistical properties of the random map  \eqref{determ+noise} can be investigated by studying the {\it annealed transfer operator} $L_{\theta}:L^{1}\rightarrow L^{1}$, which is the averaged transfer operator $L_{\xi,\theta}$ over $\xi$ with the distribution $\rho_{\theta}$, defined by 
\begin{equation}
    \label{dfn:annealed_operator}
    L_{\theta}f(A):=\langle L_{\xi, \theta}f(A)\rangle_{\xi}=\int \int_A  (L_{\xi,\theta}f)(x) dx\,  \rho_{\theta}(\xi)d\xi,
\end{equation}
remark that the inner integral depends on $\xi$.

A fixed point $f_{\theta}(x)$ of the annealed transfer operator $L_{\theta}$ satisfying
\begin{equation}
    \label{dfn:stationary_dist}
    L_{\theta}f_{\theta}(x)=f_{\theta}(x).
\end{equation}
is called a stationary distribution of the random map \eqref{determ+noise} and characterizes  (some of) the statistical properties of an ergodic random dynamical system \citep{barreira2001lectures}.
Using a stationary distribution $f_{\theta}$, we can define the spatially averaged Lyapunov exponent 
\begin{equation}
    \label{dfn:lyap_on_density_annealed}
    \langle\lambda_{\theta}\rangle=\int_{\mathbb{R}} \log|T'(x)|f_{\theta}(x)dx.
\end{equation}
Thus, when the random dynamical system is ergodic with respect to the measure $\mu_{\theta}$ whose density is $f_{\theta}(x)$ 
we have
\begin{equation}
    \label{eq:ergodicity}
    \overline{\lambda_{\theta}}=\langle\lambda_{\theta}
    \rangle
\end{equation}
for $\mu_{\theta}$-almost all $x_0$ and for almost every realization of the noise $\omega$ \citep{barreira2001lectures}. 

We herein rigorously compute the spatially averaged Lyapunov exponent $\langle\lambda_{\theta}\rangle$. Since the convergence to the equilibrium of the system ad hence its ergodicity are also proved  by our rigorous computation, we can  evaluate $\overline{\lambda_{\theta}}$ and show the existence of the multiple noise-induced transitions. 


\section{Noise-induced transitions in the Lasota-Mackey map}
\label{sec:target}
In this section we introduce the class of systems we are going to investigate. We also show the results of several non-rigorous numerical experiments. 
Beside showing the general behavior of noise induced phenomena occurring in Lasota-Mackey maps, the result of the  nonrigorous experiments we show, will help us to find the right parameters for which a multiple transition occur, and then apply to these examples our theory and computer aided estimates, proving rigorously the exixtence of the multiple transition.
\subsection{Multiple noise-induced multiple transition in the Lasota-Mackey map}
\begin{figure}[hbtp]
    \includegraphics[scale=0.35]{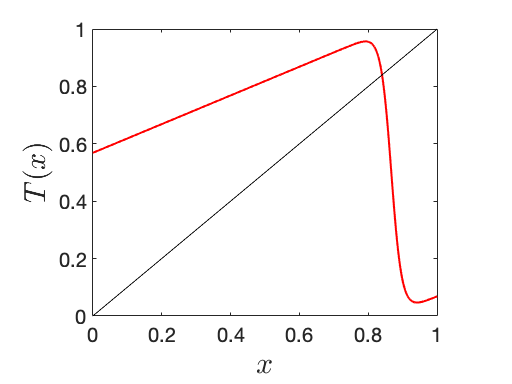}
    \caption{The deterministic Lasota-Mackey map with the parameters  $a=1/2, \beta=130, d=17/30, b=0.0018$ (red) and the identity map $y=x$ (black).}
    \label{fig:model_shape}
\end{figure}

The Lasota-Mackey map is a class of one-dimensional random map with
the deterministic term given by 

\begin{equation}
\label{eq:lasota-mackey}
    T(x)=ax+d -\frac{1}{1+e^{-\beta (ax+d-1)}}+b,
\end{equation} 
where $a,d,\beta>0$ and $b\in \mathbb{R}$\citep{aihara1990chaotic}, 
as depicted in Fig \ref{fig:model_shape}. The  stochastic term $\xi_n$, i.i.d sampled from a uniform distribution $\rho_{\theta}(x)=1_{[-\frac{\theta}{2},\frac{\theta}{2}]}$. 
When $\theta=0$ and $\beta \rightarrow \infty$, the deterministic term is given as $x_{n+1}=ax+d-1+b$, and when defined in a circle, is a classical model of neurons, so called Nagumo map \citep{nagumo1972response, lasota1987noise}. Hereafter, the parameters are fixed as $a=1/2, ~d=17/30, ~\beta=130 $\cite{lasota1987noise}. The parameter $b$ and $\theta$ are left as control parameters. In this section, we show a qualitative description of the behavior of the Lyapunov exponent of the map when varying $\theta$ and $b$. 

To get a first intuitive understanding of the behavior of the Lyapunov exponent, let us  discuss the behaviour of the Lasota-Mackey map in the {\it large noise limit}. The large noise limit is a limit in which the noise amplitude is sufficiently large to smear out whole dynamical structure.  In this limit, (i) the stationary distribution approaches to the probability distribution of the noise itself, and (ii) the trajectories almost always stay a region with $|x|\gg 1$ (see Fig. 1). With (i) and (ii), we have the following: 
  \begin{enumerate}
    \item{Spatially averaged Lyapunov exponent 
  \begin{equation}
\langle\lambda_{\infty}\rangle
    =\lim_{\theta \rightarrow \infty} \int_{-\frac{\theta}{2}}^{\frac{\theta}{2}} \log |T'(x)|\frac{1}{\theta}dx \simeq\log a <0, 
  \end{equation}
  }
\item{Temporally averaged Lyapunov exponent
  \begin{equation}
        \overline{\lambda_{\infty}} = \lim_{N\rightarrow\infty}\frac{1}{N}\sum_{i=0}^{N}\log |T'(x_i)|\simeq\log a <0.  
  \end{equation}
  }
\end{enumerate}
Therefore, starting with a Lasota-Mackey map with 
a positive Lyapunov exponent, adding a very large noise, it shows negative Lyapunov exponent. 

However, this observation does not enable us to understand noise-induced phenomena in realistic systems. As a matter of fact in the random systems we consider  we observe {\it multiple noise-induced transitions when the additive noise is small}, which cannot be discussed by the large noise limit. 

  Next, we numerically (non-rigorously) compute Lyapunov exponents as a function of $b$ and $\theta$ and have a global view in a diagram of Lyapunov exponents. In nonlinear physics, the finite-time Lyapunov exponent is often introduced as an approximation of the Lyapunov exponent, which is a function of a finite trajectory of length $N$ on the attractor. The concept of attractor in random dynamical systems is introduced similarly to those in the deterministic dynamical systems, which is called random attractor $\mathcal{A}(\omega)$\citep{fla,chekroun2011stochastic, arnold1995random}. The finite-time Lyapunov exponent on a random attractor $\mathcal{A}(\omega)$ in a random dynamical system is given by 
\begin{equation}
\label{dfn:lyap_finite_time}
    \lambda_{\theta}(N,\omega,x_{0})=\frac{1}{N}\sum_{i=0}^{N-1}\log |T'(x_{i})| ~~(x_0\in \mathcal{A}(\omega)).
\end{equation}
The temporary averaged Lyapunov exponent can be given by a long-run limit of the finite-time Lyapunov exponent. 

\begin{figure}[hbtp] 
 \centering 
 \includegraphics[scale=0.45]{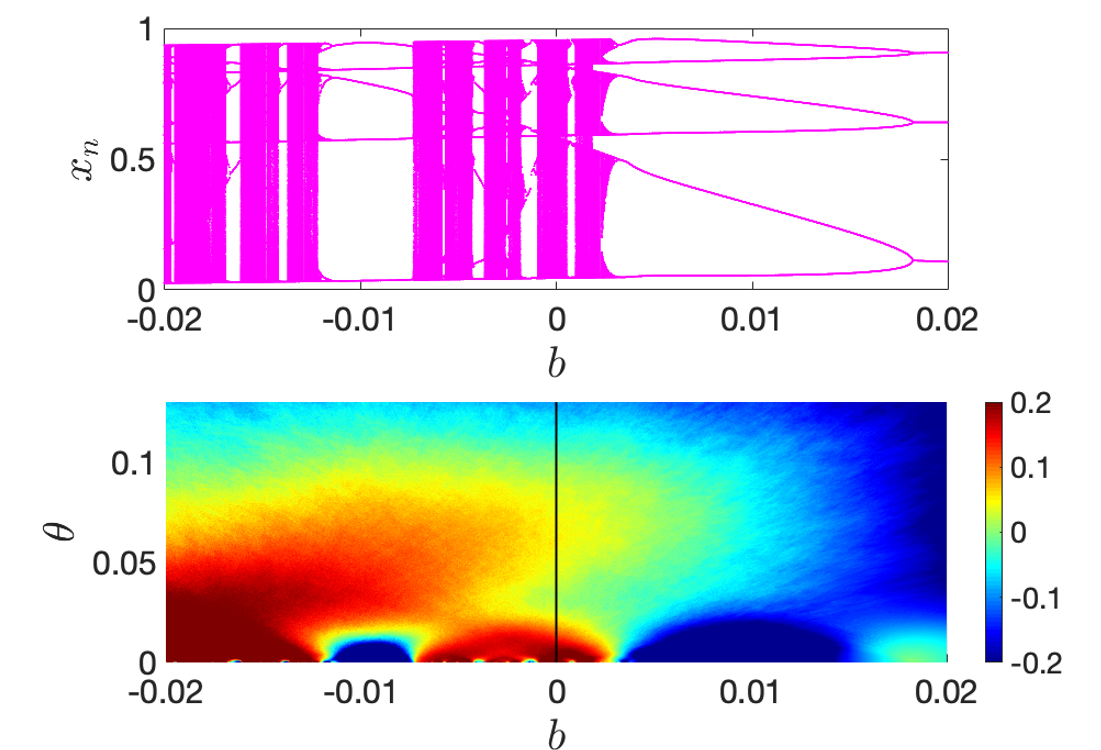}
 \caption{(top) The bifurcation diagram of the deterministic Lasota-Mackey map by changing $b \in [-0.02,0.02]$: The parameters are set to  $a=1/2, d=17/30, \beta=130$, and $\theta=0$.  (bottom) The phase diagram of the Lasota-Mackey map: The parameters are set to $a=1/2, d=17/30, \beta=130$. The Lyapunov exponent {\color{red} is} approximated by finite-time Lyapunov exponent \eqref{dfn:lyap_finite_time} in $b \in [-0.02,0.02]$ and $\theta \in [0,0.13]$. The colors correspond to  the value of Lyapunov exponents. The black line on $b=0.0018$ corresponds to the example of the parameter where a multiple transition occurs. Numerical computations are done with $10^6$ initial conditions for each $(b,\theta)$ and the finite-time Lyapunov exponent is averaged over $10^5$ time steps after $\sim 10^6$ steps of transient dynamics.}
 \label{fig:diagrams}
\end{figure}
\begin{figure}[hbtp]
 \centering 
 \includegraphics[scale=0.3]{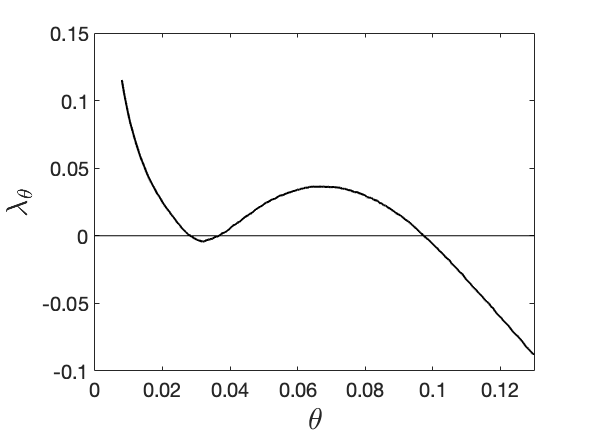} 
 \caption{The finite-time Lyapunov exponents averaged over $10^5$ time steps after $\sim 10^6$ step transient dynamics.}  
 \label{fig:lyap_only}
\end{figure}

Fig.\ref{fig:diagrams} (top)  shows the bifurcation diagram of the deterministic Lasota-Mackey map fixing $\theta=0$ and changing the shift parameter $b$. 
The finite-time Lyapunov exponents as a function of ($b, \theta$) is shown the heatmap diagram in Fig.\ref{fig:diagrams} (bottom). 
The warm color regions correspond to positive Lyapunov exponents and the cold color regions to negative Lyapunov exponents; one can see that
the warm color regions lean to the left and the multiple transitions are observed in a broad range of the parameters $b\in [-0.02,0.02]$.  
When the parameter $b = 0.0018$, we observe multiple transitions from chaos to order, and to chaos, and to order by increasing the noise amplitude $\theta$ (Fig.\ref{fig:lyap_only}).


In the next section IV, we will prove that these transitions actually occur by our rigorous computer aided estimates based on the approximation of the transfer operators.

\subsection{Regularization of stationary distributions by additive noise}

Transfer operators and stationary distributions of deterministic/random dynamical systems are typically approximated by the Ulam method (see Appendix \ref{appendix:Ulam method}).  Given a coarse-grained grid which has $K$ intervals with size $\delta=1/K$, we have a $K\times K$ probability matrix $L_{\delta,\theta}$ as an Ulam discretised annealed transfer operator $L_{\theta}$, and a $K$-dimensional probability vector $f_{\delta,\theta}$ as an Ulam discretised stationary distribution $f_{\theta}$. 

\begin{figure}[htbp]
 \centering 
 \includegraphics[scale=0.3]{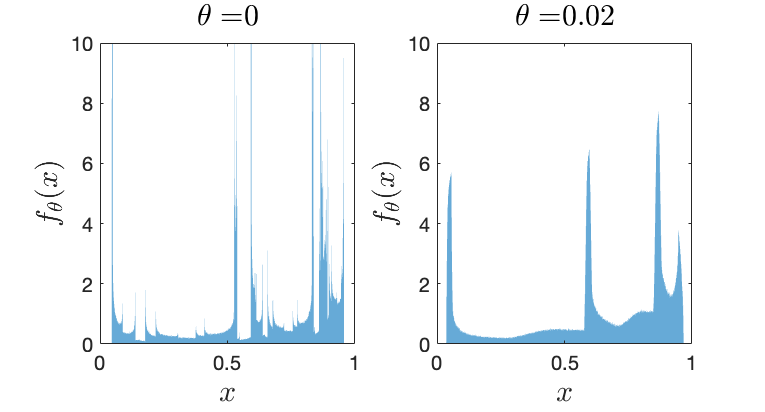}
 \caption{Non-rigorously approximated stationary distributions of the Lasota-Mackey maps ($a=1/2,~d=17/30,~\beta=130,~b=0.0018$) with $\theta=0$ (left) and with $\theta=0.02$ (right). 
 The stationary distribution is approximated by a histogram of finite trajectory with long double precision, the bin size of the histogram is $3.0\times 10^{-3}$ and the length of the trajectory is $10^{7}$, of which 90$\%$ was truncated as transients.}
 \label{fig:noise_regularized}
\end{figure}

It is difficult to obtain the rigorous error bound $\|f_{\theta}-f_{\delta,\theta}\|$ for deterministic dynamical systems and in particular for our Lasota-Mackey maps because the associated transfer operators might have a complicated  behavior, which could be unstable to perturbations as the finite dimensional approximations we need for the computation. (See \citep{froyland2000rigorous, hunt1996estimating, GN16, GNM16, galatolo2014elementary, OI} as tractable cases). Furthermore these systems may have invariant measures which are singular or the associated densities may have many sharp peaks (see Fig. \ref{fig:noise_regularized} (left)) and be supported on complicated attractors. To the contrary, it is typically easier to obtain the rigorous error bound $\|f_{\theta}-f_{\delta,\theta}\|$ for dynamical systems with additive noise, 
because the additive noise regularizes the behavior of the associated transfer operator. In our case indeed, due to the Bounded Variation noise kernel $\rho_\theta$, the associated transfer operator is  regularizing from $L^1$ to Bounded Variation (see \citep{galatolo2017existence}), and hence it is compact Markov operator on $L^1$. 
The smoothing effect induced by noise 
even at the level of stationary densities 
is illustrated in Fig \ref{fig:noise_regularized} (right). 

In next section, we approximate the stationary distribution $f_{\theta}(x)$ with rigorous error bound, 
by using the Ulam method. 

\section{Rigorous approximation of Lyapunov exponents}
\label{sec:theory}
\subsection{Approximation of stationary distribution}\label{AST}
In Galatolo {\it et al.}\citep{galatolo2017existence}  is given the algorithm that bounds the error in approximating the stationary distribution of random dynamical system \eqref{determ+noise} with the Ulam method (see Appendix \ref{appendix:Ulam method}).
The rigorous computation algorithm is established for a random dynamical system on a finite interval $[0,1]$.
Although dynamics of the Lasota-Mackey map is defined on the real line, it can be reduced to whose in a bounded interval. In fact the Lasota-Mackey maps have a compact attracting set. Let us consider the Lasota-Mackey map  (\ref{eq:lasota-mackey}) with coefficients $a=1/2, ~d=17/30, ~\beta=130,~ b=0.0018$ and recall that $|\xi_n|\leq \theta$. Let $s_{1},s_{2}$  $(s_1<s_2)$ be the critical points of  $T$. 
Let us consider the interval $I=[T(s_{2})-\theta/2,T(s_{1})+\theta/2]$.
Let $x_0$ be an initial condition and $\xi_n$ be some realization of the noise.  Since $a<1$  after a finite number $n$ of iterates  we get $x_n\in I$ and then eventually $x_i\in I$ for each $i\geq n$. The interval $I $ includes an attracting set for every random orbit of the system. By this any initial distribution of probability will be sent to a distribution supported in $I$. $I$ will hence contain the support of the stationary measure of the system. We can then consider the random dynamics restricted to $I$ and apply our techniques to compute the stationary measure and the associated Lyapunov exponents.

We explain here the general idea used in Galatolo {\it et al.} \citep{galatolo2017existence}, allowing the possibility of finding explicit error bounds between the stationary distribution $f_{\theta}$ of the random dynamical system \eqref{determ+noise} and the stationary measure $f_{\delta,\theta}$  of the Ulam discretization of the system.
We expose here a kind of simplified version of the construction used in the paper \citep{galatolo2017existence}, with the aim of showing the aspects of the system having the greatest influence on the speed and the precision of our explicit estimates: the speed of mixing of the system which determines the number of iteration required to certify the bounds, and the size of the noise, which is responsible for the regularization effect of the transfer operator.

Recall that the $L_{\theta}, L_{\delta,\theta}$ are respectively the annealed transfer operator and the Ulam discretization of the random dynamical system \eqref{determ+noise}.
Let $V\subseteq L^1$ be the set of zero average densities
\begin{equation}
  V:=\{ f\in L^1 \ s.t. \int f dx=0\}.
\end{equation}
Let $v_i:= \| L^{i}_{\delta,\theta}|_V\|$ the $L^1$ norm of the approximating transfer operator restricted to $V$.
Suppose that there exists an integer $n$ such that 
\begin{equation}
    v_{n}<\alpha<1,
\end{equation}
(hence the transfer operator contracts $V$, implying convergence to equilibrium) and for $0\leq i<n$, 
 \begin{equation}
v_{i}<C_{i}\leq 1.
 \end{equation}
Then 
\begin{equation}
\label{eq:priori_error}
\|f_{\theta}-f_{\delta,\theta}\|_{L^{1}} \leq \frac{1+2\sum_{i=0}^{n-1}C_{i}}{2(1-\alpha)}\delta {\rm Var}(\rho_{\theta}), 
\end{equation}
where $\rm Var(\cdot)$ is total variation norm (see Appendix \ref{appendix:variation}). For the proof, of this estimate see the reference \cite{galatolo2017existence} or Appendix \ref{appendix:first bound}.
We point out that in the reference \cite{galatolo2017existence} and in the code used in the present work a more complicated and sharp bound is used, but the main concepts involved are the same as we can find in this simplified version, which will be sufficient for the purposes of this section.

For this bound, we need to compute $C_{i}, \alpha$ rigorously (for the way to compute see \citep{galatolo2014elementary}).


For our system, the probability density function $\rho_{\theta}$ is a uniform distribution on $[\theta/2,\theta/2]$, thus
\begin{equation}
{\rm Var}(\rho_{\theta})=\frac{2}{\theta},
\end{equation}
and we have
\begin{equation}
\label{eq:bound_lmmap}
\|f_{\theta}-f_{\delta,\theta}\|_{L^{1}} \leq \frac{1+2\sum_{i=0}^{n-1}C_{i}}{2(1-\alpha)}\frac{2\delta}{\theta}:=E_{1}, 
\end{equation}
From this bound, we can see a small noise size $\theta$ requires a small partition size $\delta$ to have a good approximation. Furthermore, if the system is fast mixing we will get a small value for $n$ also improving the error bound.
We remark that the  stronger bound implemented in the paper \cite{galatolo2017existence} and in the code used in this work for the computation of the stationary distribution is mostly proportional to  $\delta^{2}\rm Var(\rho_{\theta})$, improving the quality of the approximations.

\subsection{Approximation of Lyapunov exponents }
Based on the rigorous error bound of stationary distribution $f_{\delta,\theta}(x)$, we obtain the rigorous error bound of the Lyapunov exponent. 
For different system different bound are required to obtain the approximation error of the Lyapunov exponent \eqref{dfn:lyap_on_density_annealed} which is defined as $\int h(x)f(x)dx$ where $h(x)=\log T'(x)$. 
This is because the observable function $h$ diverges at the critical point, and we need to estimate differently near the critical points and in other parts of the system.
Note that Lasota-Mackey maps \eqref{eq:lasota-mackey} have two critical points $s_{1},s_{2}$ (i.e., for $i=1,2$, $T'(s_{i})=0$ and $s_{1}<s_{2}$).

Let $X$ be a space that include the support of stationary distribution $f_{\theta}$, we define the approximated Lyapunov exponent of Lasota-Mackey map as
\begin{equation}
\langle\lambda_{\delta,\theta}\rangle=\int_{X\backslash B_{\epsilon}}h(x)f_{\delta,\theta}(x)dx,
\end{equation}
where $B_{\epsilon}$ is the $\epsilon$-neighborhoods of the two critical points $s_{1},s_{2}$
\begin{equation}
    B_{\epsilon}=\{[s_{1}-\epsilon,s_{1}+\epsilon],[s_{2}-\epsilon,s_{2}+\epsilon]\} ~~(\epsilon>0)
\end{equation}
Applying the $L^{1}$ norm instead of $L^{\infty}$ in $B_{\epsilon}$, we have
\begin{gather} 
\begin{split}
&\left|\langle\lambda_{\theta}\rangle-\langle\lambda_{\delta,\theta}\rangle\right|\\
    &=\left|\int_{X}h(x)f_{\theta}(x)dx-\int_{X\backslash B_{\epsilon}}h(x)f_{\delta,\theta}(x)dx\right| \\ 
    \label{eq:error_observable}
    &\leq \|h(x)\|_{L^{1}(B_{\epsilon})}\cdot \|f_{\theta}(x)\|_{L^{\infty}(B_{\epsilon})}+c\|f_{\theta}(x)-f_{\delta,\theta}(x)\|_{L^{1}}\\
\end{split} 
\end{gather}
where $c$ is constant value given as
\begin{equation}
    c=\log\left|a(1+\frac{\beta}{4})T'(s_{1}-\epsilon)\right|.
\end{equation}
Note that the bound of $\|f_{\theta}(x)-f_{\delta,\theta}(x)\|_{L^{1}}$ is given by previous section as (\ref{eq:bound_lmmap}).
We give the bound of $\|h(x)\|_{L^{1}(B_{\epsilon})}$ as 
\begin{gather} 
\begin{split}
    &\|h(x)\|_{L^{1}(B_{\epsilon})}\\
    &=\int_{s_{1}-\epsilon}^{s_{1}}|h(x)|dx+\int_{s_{1}}^{s_{1}+\epsilon}|h(x)|dx \\
    &\ \ \ \ \ +\int_{s_{2}-\epsilon}^{s_{2}}|h(x)|dx+\int_{s_{2}}^{s_{2}+\epsilon}|h(x)|dx \\
    &< m_{1}(s_{1}-\epsilon)+M_{1}(s_{1}+\epsilon)+m_{2}(s_{2}-\epsilon)+M_{2}(s_{2}+\epsilon) \\
    &:=E_{2}
\end{split} 
\end{gather}
where
\begin{gather} 
\begin{split}
M_{i}(x)&=|\eta(x)-\eta(s_i)| \left[ \log|a \sigma''(\eta(x))(\eta(x)-\eta(s_i))|-1 \right] \nonumber \\
m_{i}(x)&=|\eta(x)-\eta(s_i)| \left[ \log|a \sigma''(\eta(s_i))(\eta(x)-\eta(s_i))|-1 \right ]\nonumber
\end{split} 
\end{gather}
Also we give the bound of $\|f_{\theta}(x)\|_{L^{\infty}(B_{\epsilon})}$ as 
\begin{gather} 
\begin{split}
    \|f_{\theta}\|_{L^{\infty}} &= \|L_{\theta}f_{\theta}\|_{L^{\infty}(B_{\epsilon})} \\
    &\leq \|L_{\theta}f_{\delta,\theta}\|_{L^{\infty}(B_{\epsilon})}+\|L_{\theta}(f_{\theta}-f_{\delta,\theta})\|_{L^{\infty}(B_{\epsilon})} \\
    &\leq \|f_{\delta,\theta}\|_{L^{\infty}(B_{\epsilon})}+\theta^{-1}\|(f_{\theta}-f_{\delta,\theta})\|_{L^{1}}\\
    &\leq E_{3}+\theta^{-1}E_1,
\end{split} 
\end{gather}
where $\|f_{\delta,\theta}\|_{L^{\infty}(B_{\epsilon})}:=E_3$. Therefore, we have
\begin{gather} 
\begin{split}
\label{eq:bound_E}
\left|\langle\lambda_{\theta}\rangle-\langle\lambda_{\delta,\theta}\rangle\right|< E_{2}(E_{3}+\theta^{-1}E_1)+cE_{1}:=E
\end{split} 
\end{gather}

In sum, given rigorously computed $f_{\delta,\theta},\alpha,C_i ~(i=1,\ldots, n),n$  
with (16), (17), (18), (19), 
we obtain the certificated interval of the Lyapunov exponent
\begin{equation}
\langle\lambda_{\theta}\rangle \in (\langle\lambda_{\delta,\theta}\rangle-E,\langle\lambda_{\delta,\theta}\rangle+E). 
\label{eq:lbound}
\end{equation}

\section{Existence of multiple noise-induced transitions in Lasota-Mackey maps}
\label{sec:result}
\subsection{Rigorous computation based on stationary distribution}
We apply the computational library based on the rigorous estimation introduced previous sections for our Lasota-Mackey map. The library written in Python, Sage, and C++, can be found at the library site (see the data availability statement). 
The main algorithm works as follows:
\begin{enumerate}
\item{Given the partition size $\delta$, compute $L_{\delta,\theta}$ and $f_{\delta,\theta}$}. 
\item{Find a small $n$ and compute $\alpha,~ (C_{i})_{i=1\dots n-1}$ satisfying the condition 
$v_{n}\leq \alpha < 1, v_{i}<C_{i}\leq 1 ~(i=0\dots n-1)$} 

\item Compute the rigorous error bound $E_1$ of the approximated stationary distribution  $\|f_{\theta}-f_{\delta,\theta}\|_{L^{1}}$ 
\item Compute the approximated Lyapunov exponent $\langle\lambda_{\delta,\theta}\rangle$ and the rigorous error bound $E$.
 $\langle\lambda_{\theta}\rangle$ is in the interval $(\langle\lambda_{\delta,\theta}\rangle-E,\langle\lambda_{\delta,\theta}\rangle+E)$. 
\end{enumerate}
In step 2, we need to compute a high dimensional matrix $\{L_{\delta,\theta}^{i}\}_{i \leq n}$, that mainly contributes to the computational time $O(n\delta^{-3D})$,  where $D$ is the system dimension.
When the contraction of approximated transfer operator $L^{n}_{\delta,\theta}$ is certificated, the contraction of original annealed transfer operator with $n+1$ iteration $L^{n+1}_{\theta}$ is certificated (see Section 4 of the paper \citep{galatolo2017existence}). In the other words, we can certificate the mixing property of the system by a secondary result of the rigorous computation. 

The fact that the system is contracting zero average measures and hence mixing, by the results explained in Section 7 of the paper \citep{galatolo2017existence} imply that the Lyapunov exponent is H\"{o}lder continuous as a function of $\theta$ ($\theta>\theta_{0}$). If the system is mixing with additive noise with the amplitude $\theta_{0}$, the system with a larger fluctuation $\theta>\theta_{0}$ is also mixing. These facts support the existence of the zero-crossing points of the Lyapunov exponents when the noise-induced transition exists. 

\begin{figure}[hbtp]
 \centering 
 \includegraphics[scale=0.25]{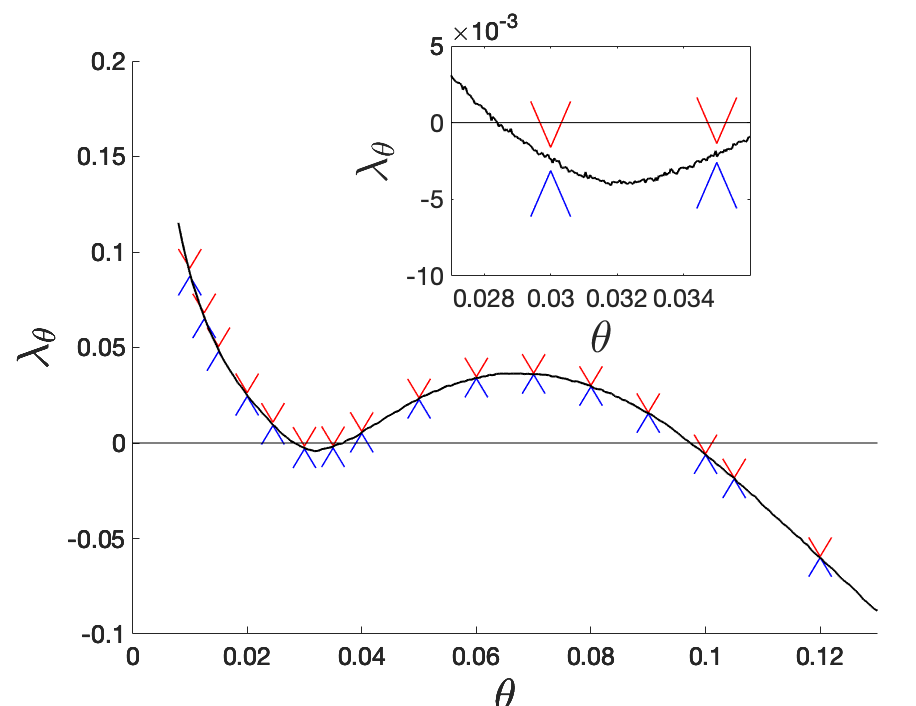} 
 \caption{The rigorously computed interval including the Lyapunov exponents:  The certificated intervals are given by red and blue arrow. The precise value of certificated interval is on the Table 1. The parameters are set to $a=1/2, d=17/30, b=0.0018, \beta=130$, and 16 sampled $\theta\in[0.7250\times 10^{-2},9.000 \times 10^{-2}]$. The black line is the non-rigorous approximation given by the finite time Lyapunov exponents as a reference. The certified bounds hold at only these specific values of $\theta$.}  
 \label{fig:result}
\end{figure}

The final result of our rigorous approximation of the certificated interval of the Lyapunov exponents is shown in Table \ref{tab:result} and Fig.\ref{fig:result}. We give the partition size and the noise amplitude
\begin{equation}
    \delta=2^{-20}, ~~\theta\in[0.01, 0.12], 
\end{equation}
with $b=0.0018$.
The algorithm automatically finds the iteration number $n$ and the contraction rate $\alpha$ to output the $L^1$ error $E'\simeq 10^{-3}$. Note that, in the implemented algorithm, we adopt the bound not as $E$ in Eq. \eqref{eq:bound_E}, but as a stronger bound $E'$ (see Sec. 3.3 in the reference \citep{galatolo2017existence}). 
The stronger bounds is given as $E'\propto \delta^2/\theta$, while the standard bound as $E\propto \delta/\theta$.
\begin{table}[hbtp]
    \centering
    \scalebox{1.0}{
    \begin{tabular}{|c|c|c|c|c|} \hline
    $\theta$  & $\delta$ & $n$ & $\alpha$ & $(\langle\lambda_{\delta,\theta}\rangle-E',\langle\lambda_{\delta,\theta}\rangle + E')$  \\ \hline \hline
    $1.000\times 10^{-2} $ & $2^{-20}$ & $48$ & $0.22$ & $[8.727, 9.158]\times 10^{-2} $\\ 
    $1.250\times 10^{-2} $ & $2^{-20}$ & $48$ & $0.17$ & $[6.490, 6.816]\times 10^{-2} $\\ 
    $1.500\times 10^{-2} $ & $2^{-20}$ & $45$ & $0.14$ & $[4.783, 5.049]\times 10^{-2} $\\ 
    $2.000\times 10^{-2} $ & $2^{-20}$ & $48$ & $0.1$ & $[2.437, 2.637]\times 10^{-2} $\\ 
    $2.450\times 10^{-2} $ & $2^{-20}$ & $46$ & $0.091$ & $[0.915, 1.087]\times 10^{-2} $\\ 
    \hline
    \rowcolor[gray]{0.8}[.8\tabcolsep]
    $3.000\times 10^{-2} $ & $2^{-20}$ & $45$ & $0.07$ & $[-3.129, -1.610]\times 10^{-3} $\\ 
    \rowcolor[gray]{0.8}[.8\tabcolsep]
    $3.500\times 10^{-2} $ & $2^{-20}$ & $46$ & $0.057$ & $[-2.595, -1.352]\times 10^{-3} $\\
    \hline
    $4.000\times 10^{-2} $ & $2^{-20}$ & $42$ & $0.056$ & $[5.036, 6.109]\times 10^{-3} $\\ 
    $5.000\times 10^{-2} $ & $2^{-20}$ & $43$ & $0.032$ & $[2.276, 2.362]\times 10^{-2} $\\ 
    $6.000\times 10^{-2} $ & $2^{-20}$ & $36$ & $0.032$ & $[3.385, 3.460]\times 10^{-2} $\\ 
    $7.000\times 10^{-2} $ & $2^{-20}$ & $39$ & $0.018$ & $[3.583, 3.651]\times 10^{-2} $\\ 
    $8.000\times 10^{-2} $ & $2^{-20}$ & $35$ & $0.017$ & $[2.951, 3.015]\times 10^{-2} $\\ 
    $9.000\times 10^{-2} $ & $2^{-20}$ & $33$ & $0.014$ & $[1.533, 1.594]\times 10^{-2} $\\ 
    \hline
    \rowcolor[gray]{0.8}[.8\tabcolsep]
    $1.000\times 10^{-1} $ & $2^{-20}$ & $32$ & $0.011$ & $[-6.198, -5.602]\times 10^{-3} $\\ 
    \rowcolor[gray]{0.8}[.8\tabcolsep]
    $1.050\times 10^{-1} $ & $2^{-20}$ & $31$ & $0.011$ & $[-1.885, -1.826]\times 10^{-2} $\\ 
    \rowcolor[gray]{0.8}[.8\tabcolsep]
    $1.200\times 10^{-1} $ & $2^{-20}$ & $29$ & $0.0087$ & $[-6.001, -5.943]\times 10^{-2} $\\ 
    \hline
    \end{tabular}
    }    \caption{The certificated interval of the Lyapunov exponent as a function of noise amplitude. It is certificated that the sign of the Lyapunov exponent is negative in gray regions, and the sign of the Lyapunov exponent is positive in white regions. The certified bounds hold at only these specific values of $\theta$.
    \label{tab:result}}
\end{table}

In the Table \ref{tab:result} white regions indicates that the upper end of the interval is negative, and the gray regions that the lower end of the interval is positive. 
We also confirm that the system (\ref{eq:lasota-mackey}) is mixing at $\theta=0.01$. This implies that the equality \eqref{eq:ergodicity} 
holds for the entire range of the parameters. 

From the above, the following theorem holds:
\begin{thm}
The Lasota-Mackey map with the parameters $a=1/2, ~b=0.0018, ~c=17/30,  ~\beta=130$ and $\rho_{\theta}(x)=\frac{1}{\theta} ~(x\in[-\theta/2,\theta/2])$, shows multiple noise-induced transitions. The sign of Lyapunov exponent changes at least three times in the interval $\theta\in[0.01, 0.12]$. 
%
\end{thm}
The mixing property also implies that the Lyapunov exponent is H\"{o}lder continuous in the entire range of the parameters. Because we have at least three change of the sign of the Lyapunov exponents, there exist at least three zero-crossing points of the Lyapunov exponents in the interval  $\theta\in[0.01, 0.12]$.

\subsection{Non-rigorous computation}

In this section we examine the reliability of the non-rigorously computed temporally averaged Lyapunov exponents, comparing it with our rigorous estimates of the spatially averaged Lyapunov exponents. 

To do that, we compute the distributions of the finite-time Lyapunov exponents \eqref{dfn:lyap_finite_time} for $20 ~000$ different finite sequences of $\omega$. As a heuristic comparison method, we adopt the three-sigma rule\citep{pukelsheim1994three}, and check whether the sample means $\pm$ three times of standard deviation includes the certificated interval.

\begin{figure}[htbp]
 \centering 
 \includegraphics[scale=0.35]{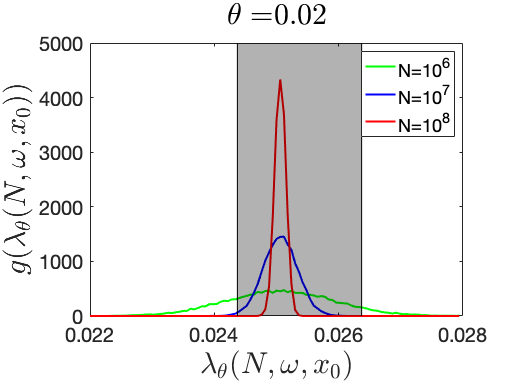}
 \includegraphics[scale=0.35]{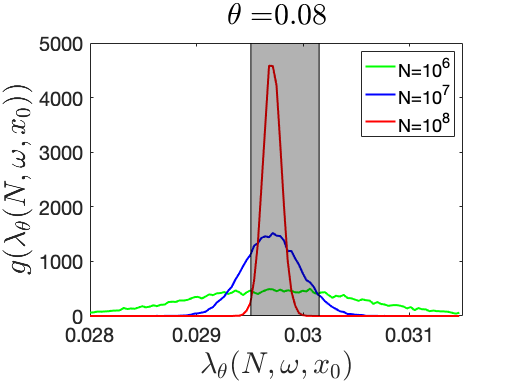}
 \caption{The certificated interval of Lyapunov exponents (gray region) and the  histogram of finite-time Lyapunov exponent for the noise amplitude $\theta=0.02$ (top) and $\theta=0.08$ (bottom).  
 Each finite-time Lyapunov exponent is given by the time average of the iteration $N=10^{6}$ (green), $N=10^{7}$ (blue), $N=10^{8}$ (red), after $9N$ steps of the  transient dynamics. 
 When $N=10^{8}$, both of the three-sigma interval are included by certificated interval.}
 \label{fig:vs}
\end{figure}

The Fig.\ref{fig:vs} exhibits the distribution $g(\lambda_{\theta}(N,\omega,x_{0}))$ of the finite-time Lyapunov exponents $\lambda_{\theta}(N,\omega,x_{0})$  
and the certificated interval of the Lyapunov exponent $[\langle\lambda_{\delta,\theta}\rangle -E',\langle\lambda_{\delta,\theta}\rangle +E']$ with the noise amplitude $\theta=0.02, 0.08$. 
Each finite-time Lyapunov exponent is given by the trajectories of length $N=10^{6}$ (green), $10^{7}$ (blue), $10^{8}$ (red), computed by the long double precision, and compute three-sigma interval which is interval [sample means $\pm$ three times of standard deviation]. 
When $N=10^{6}$, for the both $\theta=0.02$ and $0.08$, the three-sigma interval of finite-time Lyapunov exopnent don't be included by certificated interval. When $N=10^{7}$, the three-sigma interval with $\theta=0.02$ is included by certificated interval, while those with $\theta=0.08$ doesn't. 
When $N=10^{8}$, both of the three-sigma interval are included by certificated interval. Thus, in the Lasota-Mackey maps, it is suggested that the finite-time Lyapunov exponents given by the finite length time average, well-approximate the true Lyapunov exponent for a long run $N\sim 10^8$. 

Our result  then shows the reliability of the  approximations of the Lyapunov exponents by the finite-time Lyapunov exponents. 

\section{Conclusion}
\label{sec:conclusion}
We prove that a Lasota-Mackey maps shows multiple noise-induced transitions and that the sign of Lyapunov exponent in the map changes at least three times, by a rigorous computation of the certificated intervals.

The rigorous computation algorithm used here is known to be effective for a wide class of random dynamical systems with additive noise. However, we need to be cautious about whether the algorithm ends in a realistic computational time. The computational complexity of the rigorous computation is $O(n\delta^{-3D})$, where $\delta$ is the grid size of the Ulam approximation, $n$ is the mixing time of the system, and $D$ is the system dimension. The width of the certified error interval is grossly proportional to $n\theta/\delta$. Thus, in order to finish the rigorous computation in a realistic time scales, the random dynamical system of interest must be (1) not with too small noise, (2) with a short mixing time, and (3) in low-dimensional. 
In this paper, we focus only on the Lyapunov exponent as an indicator of noise-induced transition. In random dynamical systems, the sign of the Lyapunov exponent often characterizes average stability of the random pullback attractors. 
The changes of the sign of the Lyapunov exponent does not always imply bifurcations in random dynamical systems. The dynamics may show chaotic behaviour even if the Lyapunov exponent is negative \citep{sato2018dynamical}. In some cases, the Lasota-Mackey map shows a stronger orderly nature than the presented case, which indicates slow oscillatory relaxation of the density to stationary state, called noise-induced statistical periodicity \citep{sato2019noise}. It is difficult to apply our rigorous computation method to the Lasota-Mackey map showing statistical periodicity due to weak mixing. 

We also compare the results in rigorous computation and non-rigorous computation, and confirm that the non-rigorous method approximates the Lyapunov exponents of the Lasota-Mackey map with a particular parameters. By using our rigorous computation method, we may estimate the reliability of a variety of other non-rigorous approximation methods. In sum, our approach is expected to work for validating statistical properties of a broad class of nonlinear stochastic phenomena.

\vspace{5mm}

\section*{Acknowledgements} 
The authors thank to M. Monge for fruitful discussions and for advice during the implementation of this project. TC was supported by the Ministry of Education, Culture, Sports, Science and Technology through Program for Leading Graduate Schools (Hokkaido University "Ambitious Leader’s Program"). YS is supported by the Grant in Aid for Scientific Research (C) No.18K03441 and (B) No.21H01002, JSPS, Japan, and the external fellowship of London Mathematical Laboratory, UK.
SG was partially supported by the research project PRIN Project 2017S35EHN “Regular and stochastic behavior in dynamical systems” of the Italian Ministry of Education and Research (MIUR). IN was partially supported by CNPq, UFRJ, CAPES (through the programs PROEX and the CAPES-STINT project "Contemporary topics in non uniformly hyperbolic dynamics").

\section*{Author Declarations}
The authors have no conflicts to disclose. 

\section*{Data availability}
The computational libraries for the rigorous estimation of Lyapunov exponents that support the findings of this study are openly available on the web site at 
https://github.com/orkolorko/compinvmeas-python-release-noise/tree/main.  Lecture videos and associated Jupyter notebooks provided in the summer school on  computational ergodic theory at Hokkaido University are available on the web site at https://sites.google.com/view/hsi-comp-ergo-theo-2021/.

\appendix

\section{The Variation of a real function}
\label{appendix:variation}

Let $f$ be a function on the interval $X\in \mathbb{R}$, the variation of $f$ (is denoted by ${\rm Var}_{X}(f)$) is defined as follows:
\begin{equation}
    {\rm Var}_{X}(f):=\sup_{\{x_{0}<x_{1}<...<x_{k}\in X\}}\sum_{i=0}^{k-1}|f(x_{i+1})-f(x_{i})|
\end{equation}
where the supremum is taken over all possible partitions of any size $k$. If $X$ is pairwise disjoint interval, the variation is defined as sum of variation at each interval. It is known that if $f$ is smooth ${\rm Var}_{X}(f)=\|f'\|_{L^{1}}$\citep{chaosfractalnoise}. 

For example, consider the probability density function $\rho_{\theta}$ of Uniform distribution on $[\theta/2,\theta/2]$. Since $\rho_{\theta}$ varies the value by $1/\theta$ at $-\theta/2$ and $\theta/2$, that variation is given as ${\rm Var}(\rho_{\theta})=2/\theta$.
\section{The Ulam method}
\label{appendix:Ulam method}
The Ulam method enable us to approximate the transfer operator of dynamics as finite dimensional matrix. Consider the nonsingular dynamical system on $X$, then the transfer operator of the system can be defined as $L:L^{1}\rightarrow L^{1}$ . We define the Ulam discretized operator $L_{\delta}$ with associated discretizing operator $\pi_{\delta}: L^{1}\rightarrow L^{1}$:
\begin{equation}
\pi_{\delta}(g):=E(g|F_{\delta}),  \\
\end{equation}
\begin{equation}
L_{\delta}:=\pi_{\delta}L\pi_{\delta},
\end{equation}
where $F_{\delta}$ is $\sigma$-algebra associated with the partition of size $\delta$.

We consider to apply the Ulam method for the dynamical systems perturbed by additive noise. Let $\rho_{\theta}$ be probality density function of considering additive noise, where $\theta$ control fluctuation of noise. We have (annealed) transfer operator \citep{chaosfractalnoise} as:
\begin{equation}
\label{perturbedoperator}
L_{\theta}=N_{\theta}L=\rho_{\theta}*L. \\
\end{equation}
The Ulam discretization of annealed transfer operator is defined as:  
\begin{equation}
\label{def_ptf}
L_{\delta,\theta}:=\pi_{\delta}N_{\theta}\pi_{\delta}L\pi_{\delta}, 
\end{equation}
and observed that
\begin{equation}
L^{n}_{\delta,\theta}:=(\pi_{\delta}N_{\theta}\pi_{\delta}L)^{n}\pi_{\delta}
\end{equation}
taking into account that $\pi_{\delta}^{2}=\pi_{\delta}$.

\section{The bound of the $L^{1}$ error} 
\label{appendix:first bound}
We give the brief explanation about the error bounds \eqref{eq:priori_error} on the computation of the stationary distribution shown in Section \ref{AST}.  

We consider the one dimensional dynamical system with additive noise (\ref{determ+noise}) and assume that the probability distribution $\rho_{\theta}$ of the noise is in the class of bounded variation, where $\theta$ control the fluctuation of noise.

Let $L_{\theta}, L_{\delta,\theta}$ be annealed transfer operator and its Ulam approximation, let $f_{\theta},f_{\delta,\theta}$ be stationary distributions respect to $L_{\theta}, L_{\delta,\theta}$.
Suppose  $n$ is an integer such that 
$v_{n}<1$.
Then
\begin{eqnarray*}
\Vert f_{\delta ,\theta }-f_{\theta }\Vert _{L^{1}} &=&\Vert L_{\delta ,\theta
}^{n}f_{\delta ,\theta }-L_{\theta }^{n}f_{\theta }\Vert _{L^{1}} \\
&=&\Vert L_{\delta ,\theta }^{n}f_{\delta ,\theta }-L_{\delta ,\theta }^{n}f_{\theta
}+L_{\delta ,\theta }^{n}f_{\theta }-L_{\theta }^{n}f_{\theta }\Vert _{L^{1}} \\
&\leq &\Vert L_{\delta ,\theta }^{n}(f_{\delta ,\theta }-f_{\theta })\Vert
_{L^{1}}+\Vert (L_{\delta ,\theta }^{n}-L_{\theta }^{n})f_{\theta }\Vert _{L^{1}}.
\end{eqnarray*}%
Since $f_{\delta ,\theta }-f_{\theta }\in V$, and
\begin{equation*}
||L_{\delta ,\theta }^{{n}}|_{V}||_{L^{1}\rightarrow L^{1}}\leq \alpha
<1
\end{equation*}
then $\Vert (L_{\delta ,\theta }^{n} (f_{\delta ,\theta }-f_{\theta })\Vert _{L^{1}}\leq \alpha \Vert f_{\delta ,\theta }-f_{\theta } \Vert_{L^1}$ and
we have the the following
\begin{equation}
    \|f_{\theta}-f_{\delta,\theta}\|_{L^{1}}<\frac{1}{1-\alpha}\|(L_{\delta,\theta}-L_{\theta})^{n}f_{\theta}\|_{L^{1}}.
\end{equation}
$L_{\delta,\theta}-L_{\theta}$ can be decomposed as follows:   
\begin{gather} 
\begin{split}
    L_{\delta,\theta}-L_{\theta} &= \pi_{\delta}N_{\theta}\pi_{\delta}L\pi_{\delta} - N_{\theta}L\\
             &= \pi_{\delta}N_{\theta}\pi_{\delta}L\pi_{\delta}- \pi_{\delta}N_{\theta}\pi_{\delta}L\\
             &\quad +\pi_{\delta}N_{\theta}\pi_{\delta}L - \pi_{\delta}N_{\theta}L\\
             &\quad +\pi_{\delta}N_{\theta}L - N_{\theta}L.\\
             \label{decomposition}
\end{split} 
\end{gather}
By recursively decomposing $L^{n}_{\delta,\theta}-L^{n}_{\theta} $, and rearranging this, we can obtain:
\begin{gather}
\begin{split}
\label{priori}
\|(L_{\delta,\theta}^{n}-&L_{\theta}^{n})f_{\theta}\|_{L^{1}} \leq \|(\pi_{\delta}-1)f_{\theta}\|_{L^{1}}+ \sum^{N-1}_{i=0}\|L_{\delta,\theta}|^{i}_{V}\|_{L^{1}\rightarrow L^{1}} \times \\ 
&  \left(  \|N_{\theta}(\pi_{\delta}-1)Lf_{\theta}\|_{L^{1}}+\|N_{\theta}\pi_{\delta} L (\pi_{\delta}-1)f_{\theta}\|_{L^{1}} \right) \\
\end{split}
\end{gather}
For the bound $\|f_{\theta}-f_{\delta,\theta}\|_{L^{1}}$, we need to estimate the bounds of those three objects:
\begin{equation}
\label{3norms}
\|(\pi_{\delta}-1)f_{\theta}\|_{L^{1}}, \|N_{\theta}(\pi_{\delta}-1)Lf_{\theta}\|_{L^{1}}, \|N_{\theta}\pi_{\delta} L (\pi_{\delta}-1)f_{\theta}\|_{L^{1}}
\end{equation}
Those objects are made up of operators $\pi_{\delta}, N_{\theta}, L$ and invariant measure $f_{\theta}$. Note that  $\|f_{\theta}\|_{L^{1}}=1,\|L\|_{L^{1}\rightarrow L^{1}}\leq1,\|\pi_{\theta}\|_{L^{1}\rightarrow L^{1}}\leq1 $. Moreover by using total variation norm, we can obtain following bounds (for proof see Proposition 23 in the reference \citep{galatolo2017existence})
\begin{equation}
\|N_{\theta}(\pi_{\delta}-1)\|_{L^{1}\rightarrow L^{1}} \leq \frac{1}{2}\delta{\rm Var}(\rho_{\theta})
\end{equation}
\begin{equation}
\|(\pi_{\delta}-1)N_{\theta}\|_{L^{1}\rightarrow L^{1}} \leq \frac{1}{2}\delta{\rm Var}(\rho_{\theta}).
\end{equation}
From the above these two bounds and (\ref{decomposition}),(\ref{priori}) lead to the initial bound of the $L^{1}$ error 
\begin{equation}
\|f_{\theta}-f_{\delta,\theta}\|_{L^{1}} \leq \frac{1+2\sum_{i=1}^{n-1}C_{i}}{2(1-\alpha)}\delta \rm Var(\rho_{\theta}), 
\end{equation}
where $0 \leq C_{i} \leq 1$ are such that 
$v_{i}<C_{i}\leq 1.$

\bibliographystyle{plain}
\bibliography{references}

\end{document}